\documentclass[10pt]{iopart}

\usepackage{iopams}
\usepackage{graphicx}
\usepackage{dcolumn}
\usepackage{bm}
\usepackage{amssymb,amsthm}
\usepackage{lineno}
\usepackage{subfigure}
\usepackage{textcomp}
\usepackage{booktabs}
\usepackage{xcolor}

\begin{document}

\title{Superconductivity in functionalized niobium-carbide MXenes}

\author{Cem Sevik\footnote[1]{These two authors contributed equally}}
\ead{cem.sevik@uantwerpen.be}
\address{Department of Physics \& NANOlab Center of Excellence, University of Antwerp, Groenenborgerlaan 171, B-2020 Antwerp, Belgium}
\address{Department of Mechanical Engineering, Faculty of Engineering, Eskisehir Technical University, 26555 Eskisehir, Turkey}

\author{Jonas Bekaert\textmd{\dag}}
\ead{jonas.bekaert@uantwerpen.be}
\address{Department of Physics \& NANOlab Center of Excellence, University of Antwerp, Groenenborgerlaan 171, B-2020 Antwerp, Belgium}

\author{Milorad V. Milo\v{s}evi\'{c}}
\ead{milorad.milosevic@uantwerpen.be}
\address{Department of Physics \& NANOlab Center of Excellence, University of Antwerp, Groenenborgerlaan 171, B-2020 Antwerp, Belgium}
\date{\today}

\begin{abstract}
We show the effect of Cl and S functionalization on the superconducting properties of layered (bulk) and monolayer niobium carbide (Nb$_2$C) MXene crystals, based on first-principles calculations combined with Eliashberg theory. For the bulk layered Nb$_2$CCl$_2$, the calculated superconducting transition temperature ($T_c$) is in very good agreement with the recently measured value of 6 K. We show that $T_c$ is enhanced to 10 K for monolayer Nb$_2$CCl$_2$, due to an increase in the density of states at the Fermi level, and the corresponding electron-phonon coupling. We further demonstrate a feasible gate-induced enhancement of $T_c$ up to 40 K for both bulk-layered and monolayer Nb$_2$CCl$_2$ crystals. For the S-functionalized cases our calculations reveal the importance of phonon softening in understanding their superconducting properties. Finally, we predict that Nb$_3$C$_2$S$_2$ in bulk-layered and monolayer form is potentially superconducting, with a $T_c$ around 30 K. Considering that Nb$_2$C is not superconducting in pristine form, our findings promote functionalization as a pathway towards robust superconductivity in MXenes.
\end{abstract}

%

%
%
\maketitle
%
\ioptwocol
\section{Introduction}
Layered metal carbides, nitrides, and carbon-nitrides, named MXenes in the literature, have risen among the most attractive material families in recent years. Numerous studies have been published on the use of these materials, which generally have metallic properties, in numerous technological applications such as supercapacitors~\cite{super_cap}, ion batteries~\cite{battery, battery2, Yorulmaz_2020}, electromagnetic shielding~\cite{EMIS,EMIS2}, and other~\cite{app1,ReviewMX, ReviewMX2, PRM2074002}. In addition, significant progress in synthesis of MXenes has been achieved~\cite{ReviewMX3}, which created a positive feedback loop to the intensity of research on these materials. In particular, extraordinary developments have recently been reported regarding nanoengineering of functional groups covering both sides of the MXene layers, fostering custom-engineered layered MXene crystals with desired functionalities~\cite{Surface, Surface2}. 

For instance, by using substitution and elimination reactions in molten inorganic salts, Kamysbayev \textit{et al.} have synthesized high-quality layered MXene crystals that only differ by their functional group, such as Nb$_2$CCl$_2$ and Nb$_2$CS$_2$~\cite{Kamysbayev}. For these crystalline layered structures, they demonstrated the strong influence of the functional group on the electronic properties through electrical characterization. They observed a distinctive superconducting transition for Nb$_2$C\textit{T}$_2$ (with \textit{T} = Cl, S, Se) crystals with superconducting critical temperatures ($T_c$) amounting to $\sim$6, $\sim$6.5, and $\sim$4.5~K for Nb$_2$CCl$_2$, Nb$_2$CS$_2$, and Nb$_2$CSe$_2$, respectively. Wang~\textit{et al.} recently confirmed the findings for Nb$_2$CCl$_2$, obtaining a $T_c$ of 5.2~K~\cite{WANG2022101711}.

In fact, bare monolayer Nb$_2$C does not show a superconducting transition, as demonstrated experimentally in these same studies, and predicted from first-principles calculations in our prior work~\cite{D0NR03875J}. While functionalization with hydrogen strongly enhances the $T_c$ of molybdenum- and tungsten-based MXenes -- with $T_c$'s of up to 32 K predicted through first-principles calculations -- the $T_c$ of hydrogenated Nb$_2$C remains limited to 0.8--2.9 K (depending on the hydrogen positions) \cite{D2NR01939F}. 

Overall, the strong influence of the functional groups on superconductivity in MXenes is rather clear. However, the main physical mechanism responsible for inducing superconductivity upon functionalization is still not completely elucidated, with the exception of recent first-principles calculations reported for bulk Nb$_2$CCl$_2$ ~\cite{WANG2022101711} and Nb$_2$CS$_2$ in two-dimensional (2D) form~\cite{nbc2dfpt}. Therefore, in this work we have thoroughly investigated and compared the superconducting properties of the recently synthesized Nb$_2$CCl$_2$, Nb$_2$CS$_2$, and Nb$_2$CSe$_x$ crystals, in both their bulk and monolayer form.

\begin{figure}[t]
\begin{center}
                \includegraphics[width=0.8\linewidth]{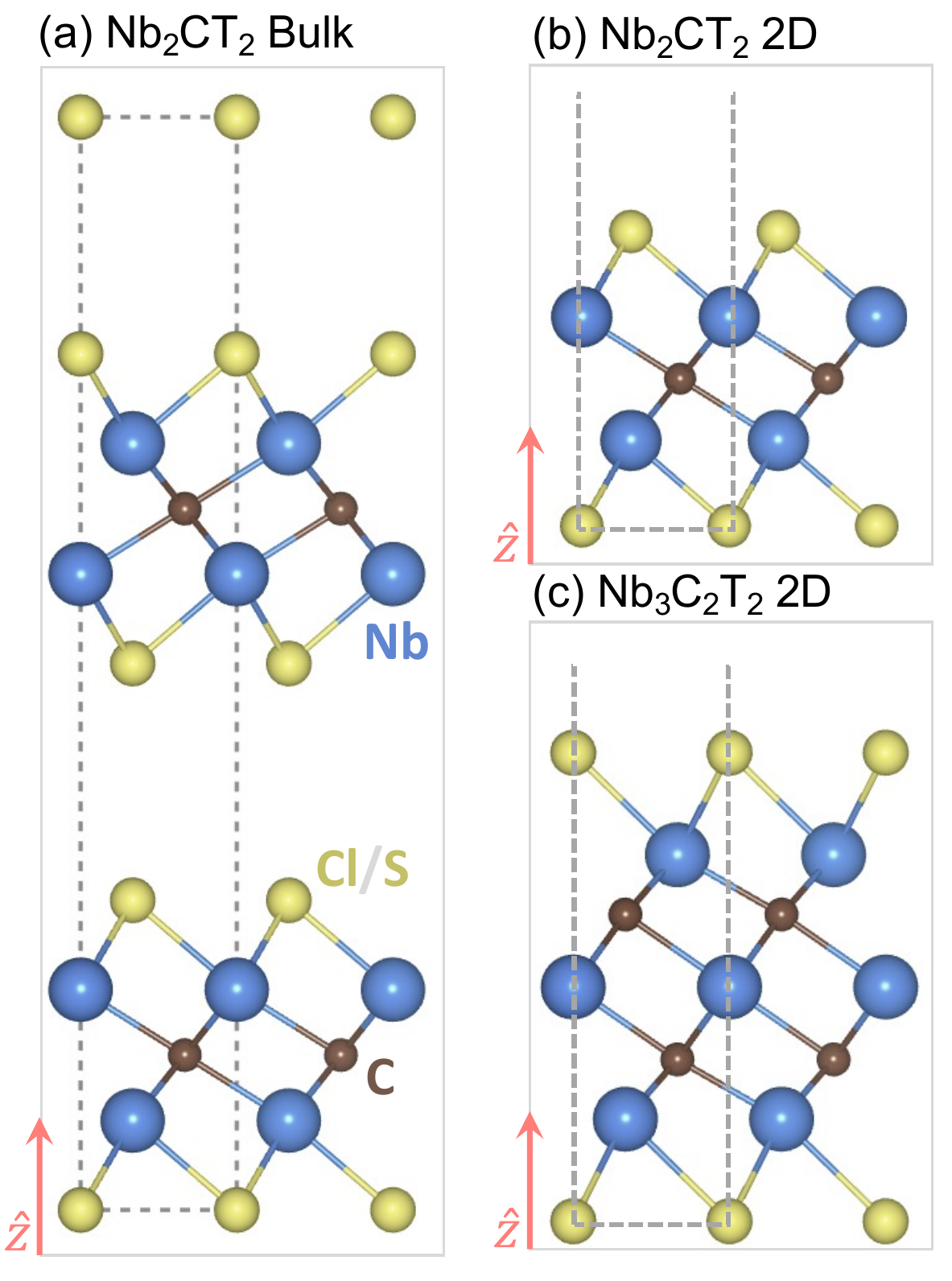}
                \caption{Crystal structures of (a) bulk-layered and (b) monolayer (2D) Nb$_2$C\textit{T}$_2$, and (c) monolayer Nb$_3$C$_{2}$\textit{T}$_2$, with \textit{T} = Cl and S. \label{fig1}}
\end{center}
\end{figure}

In contrast with the experimental results, our calculations indicate the absence of superconductivity in Se-based MXenes. Interestingly, the stoichiometry of these crystals in the experiment~\cite{Kamysbayev} deviates significantly from the ideal unit formula Nb$_2$CSe$_2$ considered in our calculations. Therefore, we will focus here mostly on the potential of functionalization with chlorine and sulfur to induce superconductivity in niobium-carbide MXenes, considering Nb$_2$C\textit{T}$_2$ (with \textit{T} = Cl and S) in bulk-layered and monolayer form, as well as Nb$_3$C$_2$\textit{T}$_2$ (with \textit{T} = Cl and S), depicted in Fig.~\ref{fig1}. 

\section{Methodology}
The calculations were performed using the density functional theory (DFT), as implemented within the ABINIT code~\cite{Gonze2020, Gonze2016}. The Perdew–Burke–Ernzerhof (PBE) type ~\cite{PBE} Hartwigsen–Goedecker– Hutter (HGH) pseudopotentials~\cite{HGH} are adopted for this purpose. The valence electron configuration of the used pseudopotentials for Nb, C, Cl, and S are $4s^{2} 4p^{6} 4d^{4} 5s^{1}$, $2s^{2} 2p^{2}$, $3s^{2} 3p^{5}$, and $3s^{2} 3p^{4}$, respectively. For all the calculations, the energy cutoff value of 50 Ha for the plane-wave basis was used. The $k$-point grids of 24$\times$24$\times$4 and 24$\times$24$\times$1 are employed for the bulk and 2D MXene crystals, respectively. The crystal structures were relaxed so all force components were below $10^{-6}$ Ha/bohr for each atom. The used vacuum space to model the 2D structures was at least 15 \AA. 

To calculate phonon dispersions and the electron-phonon (\textit{e-ph}) coupling we used density functional perturbation theory (DFPT) as implemented in ABINIT~\cite{DFPT}, using 8$\times$8$\times$1 and 12$\times$12$\times$1 phononic $q$-point grids. For the smearing of the electronic occupations around the Fermi level we used the Methfessel-Paxton method. To characterize the superconducting state we then relied on isotropic Migdal-Eliashberg theory, a quantitatively accurate extension to the Bardeen–Cooper–Schrieffer (BCS) theory for phonon-mediated superconductivity~\cite{Eliashberg1, Eliashberg2, Eliashberg3}. We evaluated the superconducting $T_c$'s using the Allen–Dynes formula \cite{PhysRev.167.331, PhysRevB.12.905, ALLEN19831}. Here, the average screened Coulomb repulsion between Cooper-pair electrons ($\mu^*$) is determined from the comparison to the available experimental measurements~\cite{Kamysbayev}, within the range of expected values for transition metal-based compounds~\cite{Grimvall}.

\section{Results}

\begin{table}[b]
\centering
\caption{Calculated and experimental structural parameters of the considered MXene crystals. TW, EXPI, and EXPII correspond to the results of this work, experimental values reported in Ref.~\cite{Kamysbayev}, and experimental values reported in Ref.~\cite{WANG2022101711}, respectively.}
\label{table1}
\footnotesize
\begin{tabular}{lcccc}
\br
MXene & Ref. & Symmetry & $a_0$ (\AA) & $c_0$ (\AA) \\\hline
\mr
Nb$_2$CCl$_2$ & TW & $P6_3/mmc$ & 3.353 & 19.947 \\
Nb$_2$CCl$_2$ & EXPI & $P6_3/mmc$ & 3.311 & 17.656 \\
Nb$_2$CCl$_2$ & EXPII & $P6_3/mmc$ & 3.162 & 17.655 \\\hline
Nb$_2$CS$_2$ & TW & $P6_3/mmc$ & 3.281 & 20.018 \\
Nb$_2$CS$_2$ & EXPI & $P6_3/mmc$ & 3.265 & 18.388 \\\hline
Nb$_2$CSe$_2$ & TW & $P6_3/mmc$ & 3.326 & 21.116 \\
Nb$_2$CSe$_2$ & EXPI & $P6_3/mmc$ & 3.282 & 23.296 \\
\br
\end{tabular}
\end{table}
\normalsize
In order to investigate the layered crystals as synthesized by Kamysbayev \textit{et al.}, we have first studied the structural properties, with the reported experimental structures as the starting point. Our results for the in-plane lattice parameters are in very good agreement with the experimental values, as seen in Table \ref{table1}. However, the deviation for the out-of-plane lattice constant, i.e. in the direction along which the MXene layers are stacked, is around 10\% for all the calculated layered MXene crystals. Therefore, the van der Waals (vdW) interaction as proposed by Grimme, based on the addition of a semi-empirical dispersion potential (DFT-D2 and DFT-D3) \cite{Grimme2006,Grimme2010}, as well as the Becke-Jonhson method \cite{Becke2006}, were tested, and improved agreement with the experimental values was reached. However, all the phonon dispersion calculation tests with inclusion of vdW interaction resulted in imaginary frequencies, which hampered the calculation of the \textit{e-ph} coupling and the superconducting properties. Therefore, we proceeded without inclusion of vdW corrections in the calculations presented here.

\begin{figure}[t]
                \includegraphics[width=\linewidth]{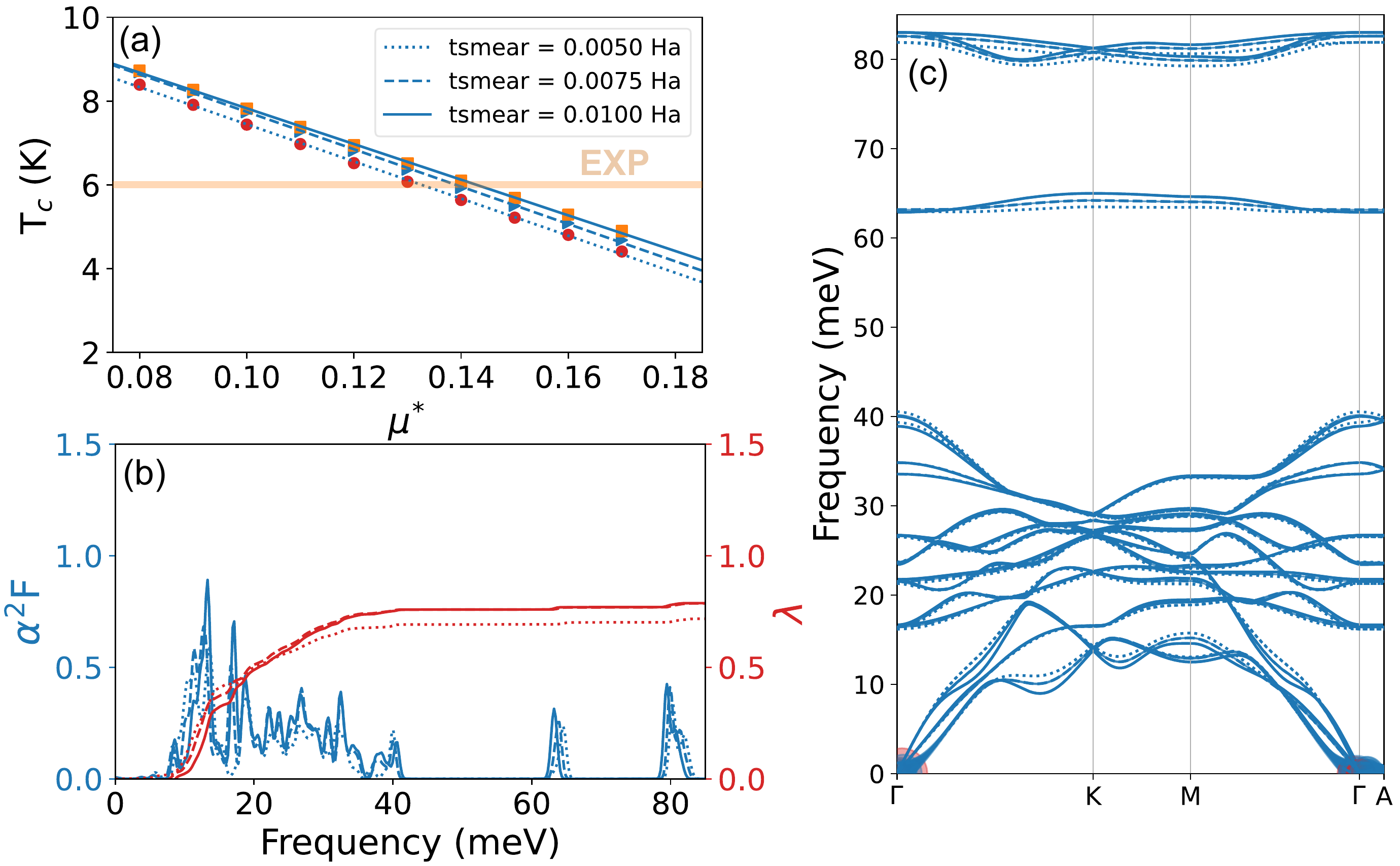}
                \caption{Superconducting properties of bulk-layered Nb$_2$CCl$_2$. (a) The dependence of the superconducting transition temperature, $T_c$, on the Coulomb pseudopotential $\mu^*$. (b) The Eliashberg function, $\alpha^2F$, and \textit{e-ph} constant, $\lambda$. (c) The phonon dispersion along with the \textit{e-ph} coupling strength, indicated by the size of the dots (red = 50$\times$blue). The solid, dashed and dotted lines in all panels represent results obtained with different electronic smearing values, \textit{tsmear}.\label{fig2}}
\end{figure}

It is a well-known fact that parameters such as the Coulomb pseudopotential ($\mu^*$), describing the effective electron-electron repulsion within the Cooper pairs, and the electronic smearing factor used in the DFPT calculations (\textit{tsmear}), may have a strong influence on the superconducting properties. Therefore, we performed a systematic analysis of the influence of these values. For Nb$_2$CCl$_2$, we found the system to be dynamically stable, with hardly any effect of the \textit{tsmear} value on the the electronic density of states (DOS) around the Fermi level (4.07, 3.51, and 3.39 states/eV per unit cell, for \textit{tsmear} = 0.0100 Ha, 0.0075 Ha, and 0.0050 Ha respectively), the phonon dispersion, or the corresponding \textit{e-ph} coupling. Therefore, the found superconducting $T_c$ values are nearly identical for all these cases, as seen in Fig.~\ref{fig2}(a). The obtained $T_c$'s are furthermore in good agreement with the experimentally measured value of $\sim$~6.0~K, for a $\mu^*$ around 0.13, which is precisely the expected value for a transition metal-based superconductor \cite{Grimvall}, see Fig.~\ref{fig2}(b). The phonon dispersion shown in Fig.~\ref{fig2}(c) clearly shows the stability of the material, even for the lowest \textit{tsmear} values. The dominant contribution of the acoustic and low-frequency optical modes on the \textit{e-ph} interaction and the resulting $T_c$ value is clearly visible. The agreement between the experiment and the presented isotropic Eliashberg calculations is very good, clearly demonstrating the phonon-mediated nature of the observed superconductivity in Nb$_2$CCl$_2$. 

\begin{figure}[t]
                \includegraphics[width=\linewidth]{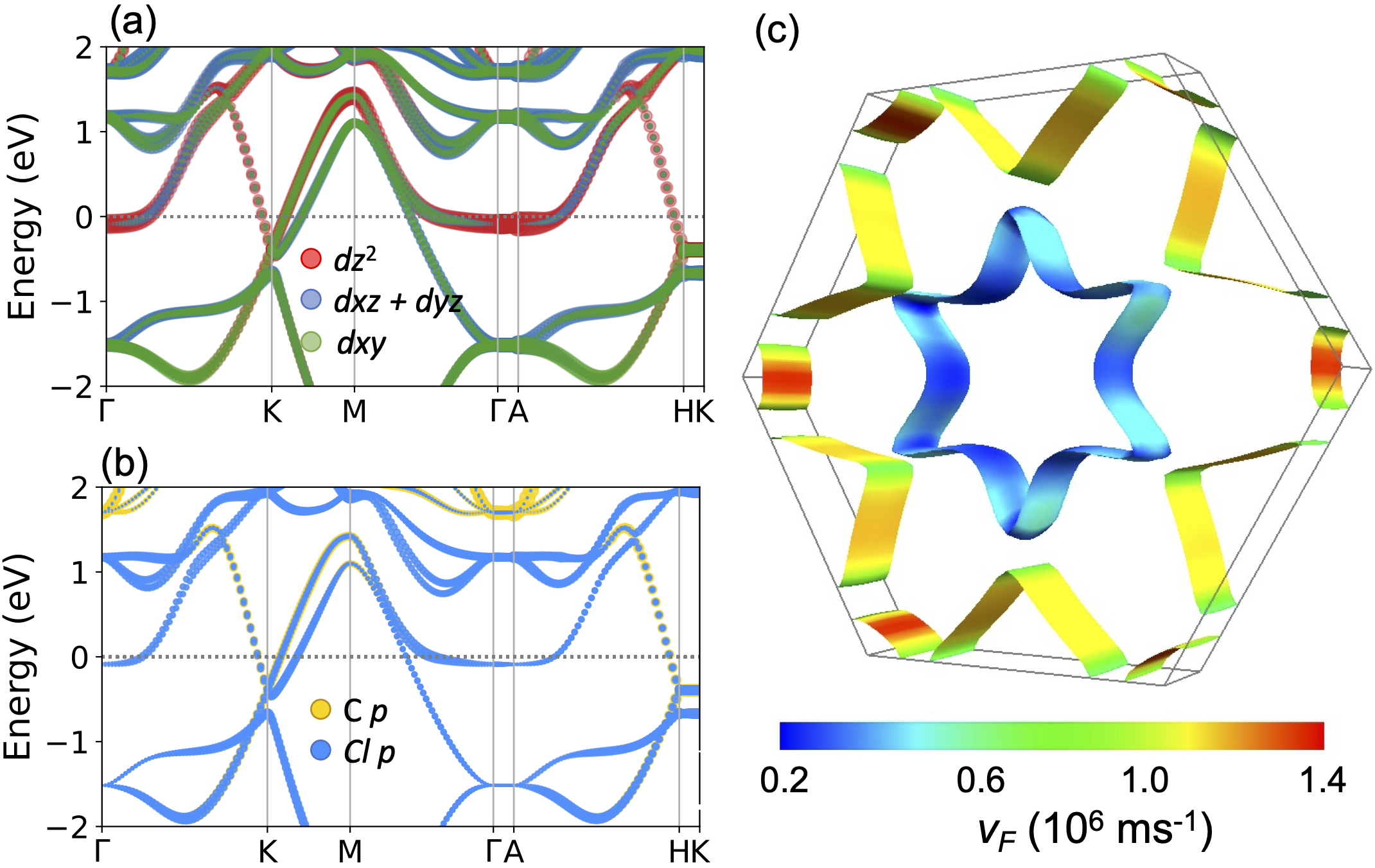}
                \caption{Electronic properties of bulk-layered Nb$_2$CCl$_2$. (a)-(b) Angular momentum-resolved electronic band structures, with the Fermi level at zero. Here, the red, blue, green, yellow and blue solid circles represent the Nb $d_{z^{2}}$, Nb $d_{xz}+d_{yz}$, Nb $d_{xy}$, C $p$, and Cl $p$ contribution respectively. (c) The Fermi surface, together with the Fermi velocities. \label{fig3}}
\end{figure}

To further characterize the origin of the superconducting state in Nb$_2$CCl$_2$, we also investigated the electronic structure. Fig.~\ref{fig3}(a--b) shows the angular momentum-resolved electronic band structure of bulk Nb$_{2}$CCl$_{2}$. The Nb $d_{z^2}$ and Nb $d_{xy}$ orbitals dominate the states near the Fermi level. The $d_{xz}+d_{yz}$ state also contributes notably in the vicinity of the $\Gamma$ and K high-symmetry points. There are also more limited contributions of C and Cl $p$ states, as shown in Fig.~\ref{fig3}(b). The resulting Fermi surface, shown in Fig.~\ref{fig3}(c) along with the Fermi velocities, consists of three distinct types of sheets: (i) a hexagonal sheet centered around $\Gamma$, stemming from a mixture of Nb $d$ states ($d_{z^2}$ mainly), (ii) six quasi-circular sheets around the K points consisting of a mixture of Nb $d_{z^2}$ and $d_{xy}$ states, and (iii) six rhombus-shaped sheets centered around the M points, mainly due to Nb $d_{xy}$ states. The $\Gamma$-centered sheet has relatively low Fermi velocities, while the sheets centered around the K points harbor the highest Fermi velocities. 

At this point we recall that entirely two-dimensional counterparts of these bulk layered MXenes can also be experimentally fabricated. Our calculations show the layer-by-layer separation energies of these crystals to be as low as 1~meV/atom -- as shown in the Supplementary Material -- especially due to their surface functional groups. Therefore, we also analyzed the properties of two-dimensional Nb$_2$CCl$_2$ crystals depicted in Fig.~\ref{fig1}(b). Our calculations show the superconducting properties of bulk and monolayer cases to be rather similar. As shown in the Supplementary Material, the calculated vibrational properties and \textit{e-ph} coupling values in a monolayer are almost the same as the ones obtained for the bulk-layered crystal. The total \textit{e-ph} coupling constant $\lambda$ amounts to 0.9, which is slightly higher than the one obtained for the bulk case (0.8). As a result of slight increase in both $\lambda$ and the electron density of states at the Fermi level, the calculated $T_c$ value of monolayer Nb$_2$CCl$_2$ is enhanced to 10~K (using the same $\mu^*=0.13$), compared to 6 K for the bulk-layered case. 

\begin{figure}[t]
                \includegraphics[width=\linewidth]{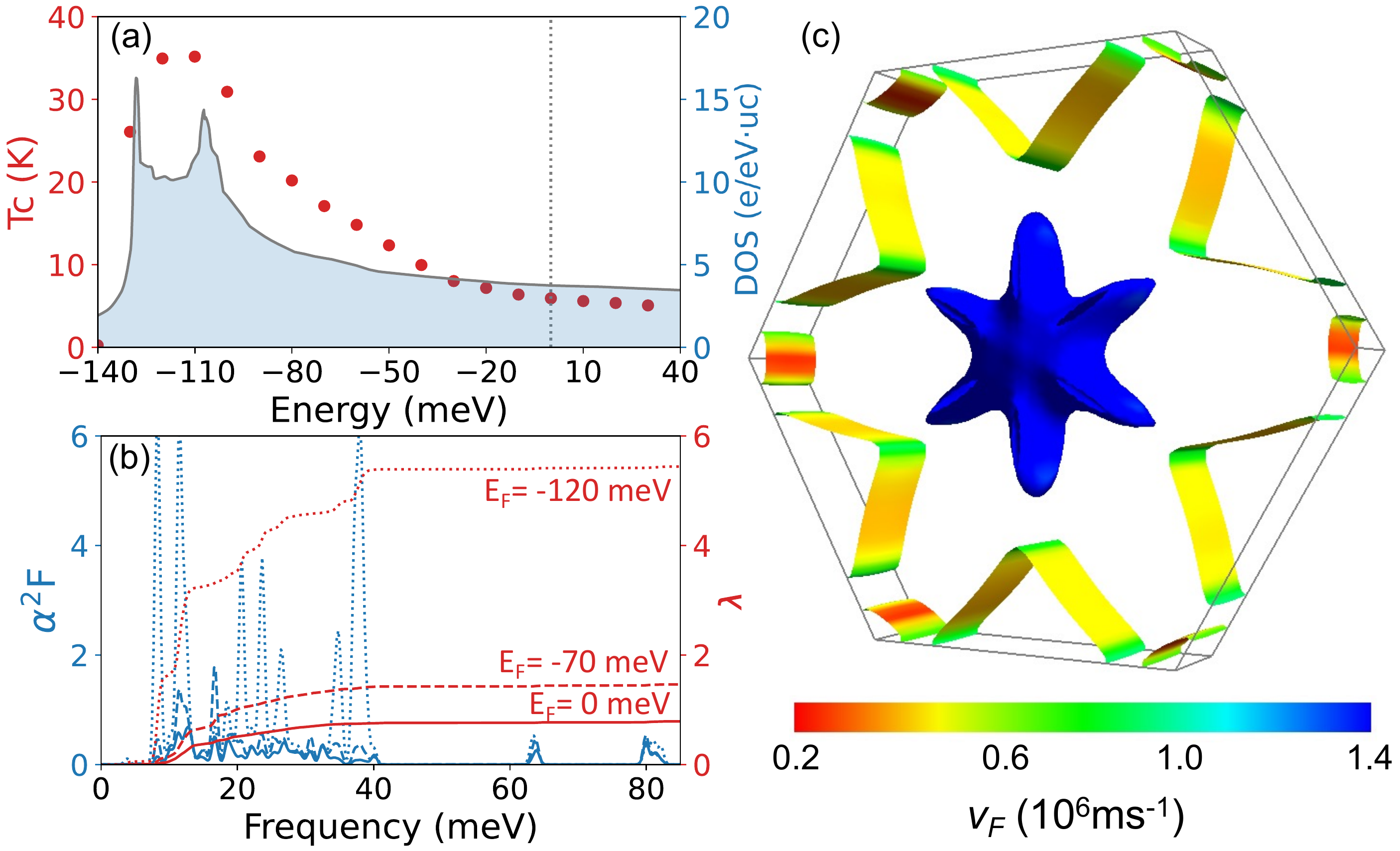}
                \caption{The effect of gating on the superconducting properties of Nb$_2$CCl$_2$. (a) $T_c$ along with the electronic density of states, as a function of the electronic energy level. (b) Eliashberg function, $\alpha^2F$, and the electron-phonon coupling constant, $\lambda$, for different values of the gating-shifted Fermi level. (c) Fermi surface, along with Fermi velocities, for the Fermi level shifted by gating to $\sim$80 meV below the intrinsic value. \label{fig4}} 
\end{figure}

Both the bulk-layered and the monolayer forms of Nb$_2$CCl$_2$ harbor a distinctly flat electronic dispersion around the $\Gamma$ and A points, right below the Fermi level (see Fig.~\ref{fig3}(a--b), and the Supplementary Material, respectively). This evokes the possibility of tailoring the superconducting properties of this material using an applied gate voltage. Therefore, we performed electron-phonon coupling and $T_c$ calculations for correspondingly shifted Fermi level values, considering both electron- and hole-type gating. Fig.~\ref{fig4}(a) shows the obtained $T_c$ values (using $\mu^*= 0.13$) of bulk-layered Nb$_{2}$CCl$_{2}$ as a function of the Fermi level shifted by gating. The $T_c$ increases up to 35 K with the shift of Fermi level coinciding with the flat dispersion. As seen in Fig.~\ref{fig4}(b), the increase in the density of states results in enhanced \textit{e-ph} coupling to both acoustic and optical vibrations within the 0--40 meV vibrational energy range. This enhanced interaction arises once the Fermi level is shifted to $\sim$ 80 meV below the original one, and paves the way for $T_c$ above 20 K. The density of states enhancement is mainly provided by the increase in surface area of the predicted six-pointed star-shaped Fermi sheet centered around $\Gamma$, as seen in Fig.~\ref{fig4}(c). In line with our findings for the intrinsic case, a similar Fermi shift boosts the $T_c$ of monolayer Nb$_2$CCl$_2$ up to 40 K (see the Supplementary Material). Therefore, these results convincingly demonstrate the prospect of using gate voltages to significantly engineer the $T_c$ of this functionalized MXene. Moreover, this is one further advantage brought by functionalization, as superconductivity in pure Nb$_2$C did not exhibit sensitivity to gating~\cite{D2NR01939F}. 

\begin{figure}[t]
                \includegraphics[width=\linewidth]{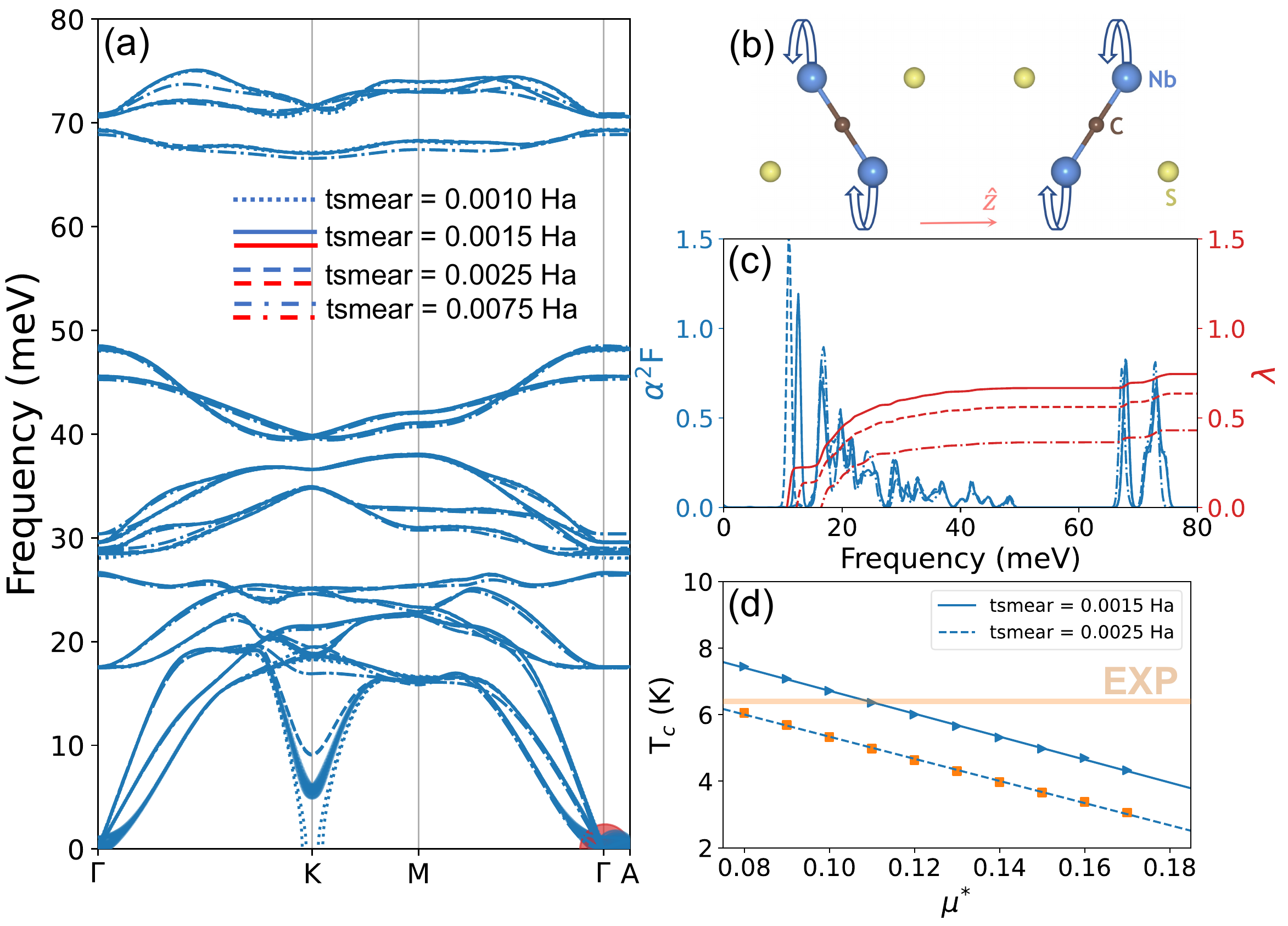}
                \caption{Superconducting properties of bulk-layered Nb$_2$CS$_2$. (a) The phonon dispersion, where the size of the colored circles indicates the strength of the \textit{e-ph} coupling (red = 50$\times$blue). (b) Atomic displacements corresponding to the soft phonon mode. (c) The Eliashberg function, $\alpha^2F$, and \textit{e-ph} coupling constant, $\lambda$. (d) The dependence of $T_c$ on the Coulomb pseudopotential $\mu^*$. The solid, dashed and dotted lines in all panels represent results obtained with different electronic smearing values, \textit{tsmear}. \label{fig5}}
\end{figure}

Unlike Nb$_2$CCl$_2$, the calculated \textit{e-ph} coupling for the S-functionalized crystal is highly sensitive to the used electronic smearing value. Computational parameters similar to the ones used for Nb$_2$CCl$_2$ (\textit{tsmear} = 0.075 Ha and $\mu^*$ = 0.13) result in a $T_c$ value of 2.0~K, whereas the experimentally measured value is 6.5 K, as indicated in Fig.~\ref{fig5}(d). Lowering the smearing value has a significant effect on the \textit{e-ph} coupling of the acoustic modes, which soften around high-symmetry point K. This phonon softening corresponds to opposite circular motion of the Nb atoms within the same layer, as shown in Fig.~\ref{fig5}(b). This results in a strong increase in $\lambda$, see Fig.~\ref{fig5}(c). Ultimately, the structure becomes unstable at smearing values around 0.0010 Ha. The other phonon branch-resolved $\lambda$ values remain nearly unchanged for all the used smearing values. The calculated \textit{e-ph} constant and $T_c$ are therefore highly sensitive to the used smearing value. A reasonable agreement with the experimental value is obtained for \textit{tsmear} around 0.0015 Ha and $\mu^*$ = 0.11, as shown in Fig.~\ref{fig5}(d). The full effect of this phonon softening, and a potential crystal structure reconstruction, is worthy of further investigation, with the inclusion of anharmonicity in the phonon spectrum \cite{PhysRevLett.125.106101}. 

\begin{figure}[t]
                \includegraphics[width=\linewidth]{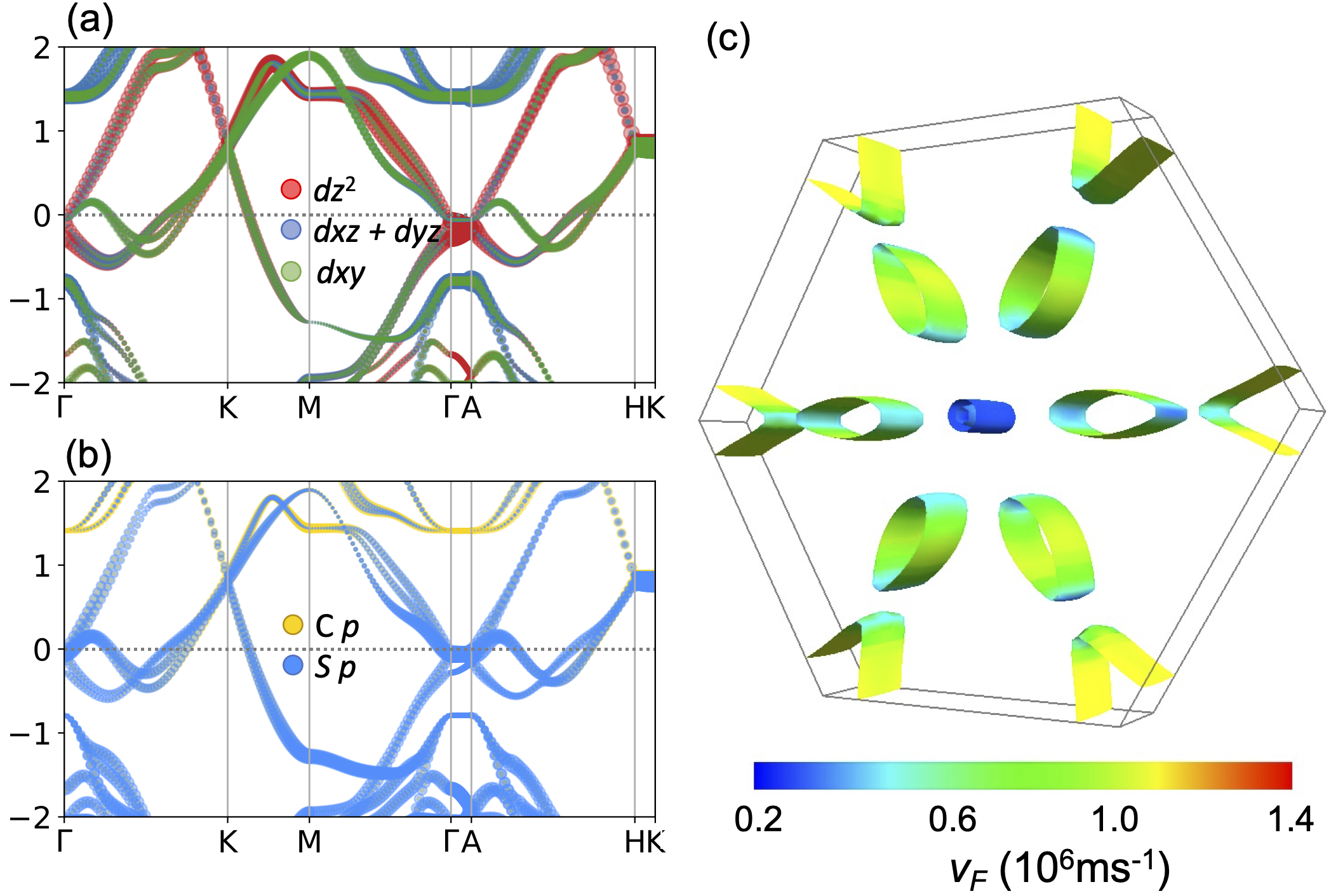}
                \caption{Electronic properties of bulk-layered Nb$_2$CS$_2$. (a)-(b) The angular momentum-resolved electronic band structures, with the Fermi level at zero. Here, the red, blue, green, yellow and blue solid circles represent the Nb $d_{z^{2}}$, Nb $d_{xz}+d_{yz}$, Nb $d_{xy}$, C $p$, and S $p$ contribution respectively. (c) Fermi surface together with the Fermi velocities. \label{fig6}}
\end{figure}

The calculated angular momentum-resolved electronic band structure of bulk-layered Nb$_{2}$CS$_{2}$ is shown in Fig.~\ref{fig6} (a) and (b). The dominant states around the Fermi level are Nb $d_{z^2}$ and $d_{xy}$, similarly to the Cl-functionalized case. However, in contrast to the Cl case, there is strong hybridization between the Nb $d$ and S $p$ states. The corresponding bands form four types of Fermi sheets, as shown in Fig.~\ref{fig6}(c). There are two $\Gamma$-centered nested cylindrical sheets, six elliptic sheets along the $\Gamma$-K path, and six rhombus-shaped sheets centered around the K points, which possess relatively high Fermi velocities compared to the other sheets. Consequently, this analysis clearly shows that the additional valence electron of Cl compared to S significantly affects the band structure around the Fermi level. Therefore, the electronic nature of superconductivity in these two compounds is distinctly different. 

\begin{figure}[t]
                \includegraphics[width=\linewidth]{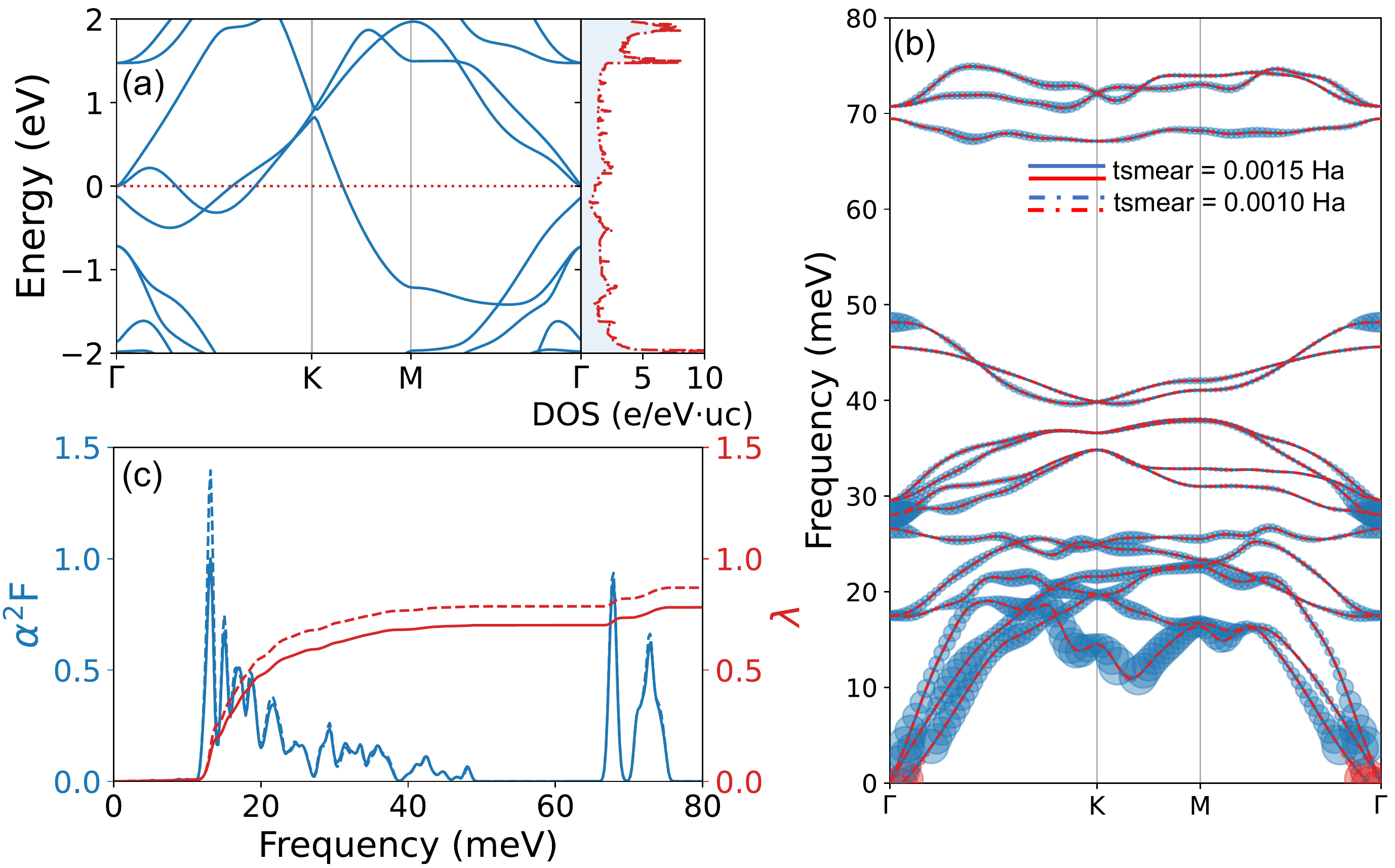}
                \caption{Electronic and superconducting properties of monolayer Nb$_2$CS$_2$. (a) Electronic band structure, with the Fermi level at zero, and density of states. (b) Phonon dispersion along with the \textit{e-ph} coupling strength. The size of colored circles shows the strength of the \textit{e-ph} coupling (red = 50$\times$blue). (c) Eliashberg function, $\alpha^2F$, and \textit{e-ph} coupling constant, $\lambda$. Here, the dashed and solid lines represent the results obtained with \textit{tsmear} values of 0.0010 and 0.0015 Ha, respectively. \label{fig7}}
\end{figure}

In line with the Cl-functionalized case, the electronic properties of monolayer Nb$_2$CS$_2$ are very similar to the bulk-layered case -- see Fig.~\ref{fig7}(a). These results clearly reveal that the chemical bonding within the layers is nearly unaffected by the stacking order, for both considered functionalization types. However, in the S-functionalized case, the vibrational properties differ notably between the bulk-layered and monolayer cases, in that the marked phonon softening found in the former is suppressed in the latter -- as shown in Fig.~\ref{fig7}(b). Other features of the vibrational spectrum are nearly identical. Only a slight softening near the high symmetry point K is present in the monolayer case, even for \textit{tsmear} values as low as 0.0010 Ha. This indicates that the monolayer form of Nb$_2$CS$_2$ is more dynamically stable than the bulk-layered one. The calculated Eliashberg function and integrated electron-phonon coupling constant, depicted in Fig.~\ref{fig7}(c), are slightly different from the bulk case. In particular, the strong coupling of the acoustic modes up to 20 meV results in enhanced $T_c$ values, 10 K and 12 K for \textit{tsmear} = 0.0015 and 0.0010 Ha, respectively. These $T_c$ values are almost twice higher than those found for the bulk case.  

In addition to the Nb$_2$CCl$_2$ and Nb$_2$CS$_2$ structures with experimentally measured Nb/surface-atom ratio close to 1, Nb$_2$CSe -- having the same space symmetry, but with a Nb/Se ratio close to 2 -- was also reported as superconducting with a $T_c$ of 4.5 K~\cite{Kamysbayev}. Hence, we have investigated Nb$_2$CSe$_2$ using the same computational approach. As shown in the Supplementary Material, the Nb$_2$CSe$_2$ crystal is dynamically stable, but it does not show superconductivity due to the weak \textit{e-ph} coupling. Here, the off-stoichiometric nature of the experimentally obtained Nb$_2$CSe crystal, which may be a compound high in Se-vacancies (potentially an ordered vacancy compound), is expected to be responsible for this apparent discrepancy between theory and experiment.

\begin{figure}[t]
                \includegraphics[width=\linewidth]{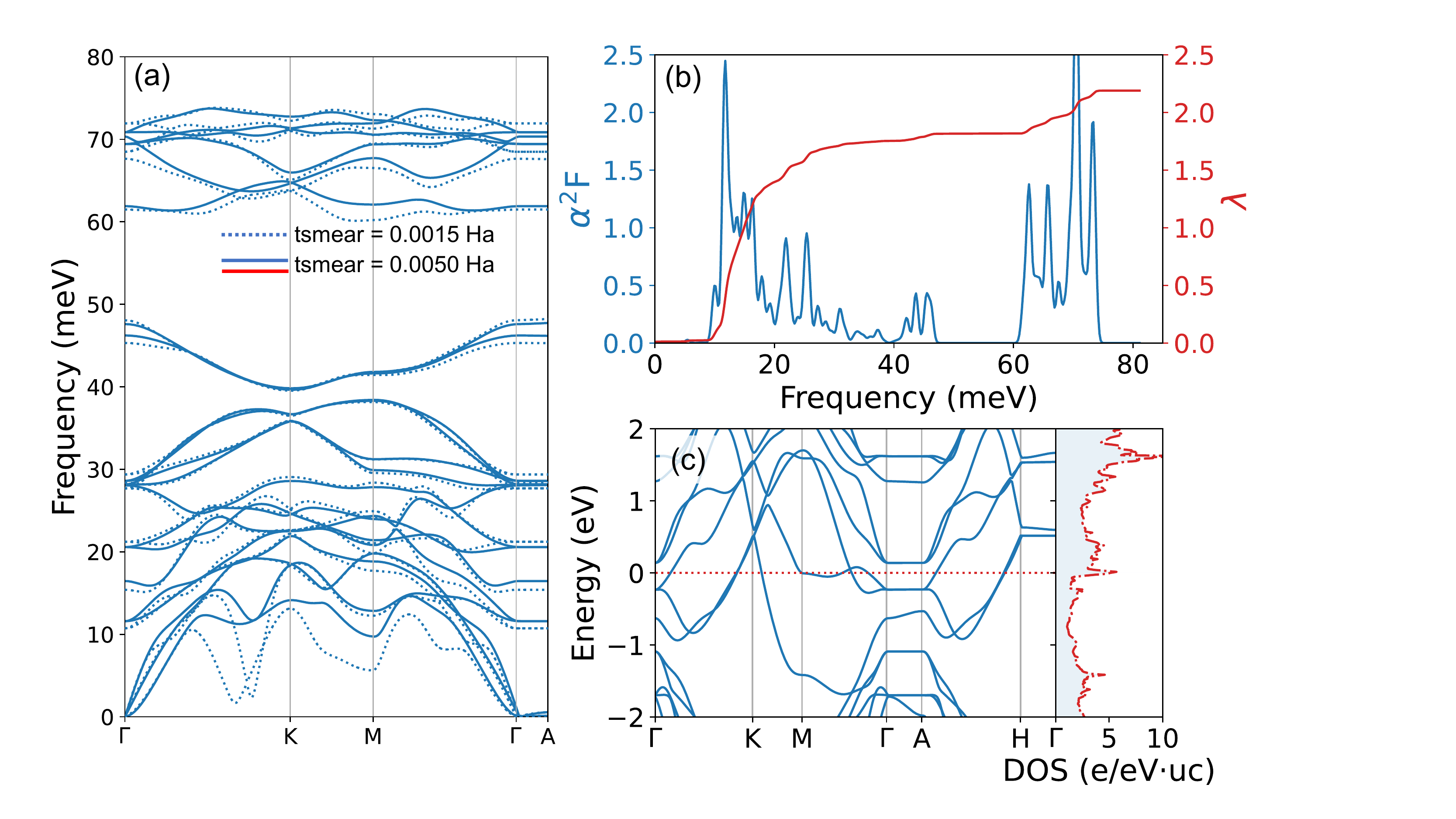}
                \caption{Electronic and superconducting properties of bulk-layered Nb$_3$C$_2$S$_2$. (a) Phonon dispersion, where the dashed and solid lines represent the results obtained with \textit{tsmear} values of 0.0015 and 0.0050 Ha, respectively. (b) Eliashberg function, $\alpha^2F$, and \textit{e-ph} coupling constant, $\lambda$. (c) Electronic band structure, with the Fermi level at zero, and the corresponding density of states.\label{fig8}} 
\end{figure}

In order to shed more light on the potential of niobium-carbide MXenes as superconductors, we also considered Nb$_3$C$_2$\textit{T}$_2$ crystals, as depicted in Fig.~\ref{fig1}(c). In the Supplementary Material we show that pristine Nb$_3$C$_2$ in 2D form is dynamically stable, but possesses only limited superconducting capabilities, with an estimated $T_c$ of merely 1 K. Due to computational limitations related to larger supercells, we only considered the stacking order with space group symmetry $P\bar{3}m1$ for the bulk functionalized structures. This corresponds to AB layer stacking resulting from a single layer in the computational unit cell. The Cl-functionalized case was found to be dynamically unstable, as shown in the Supplementary Material. On the other hand, the calculated phonon dispersion of the S-functionalized case indicates general dynamical stability, albeit in presence of a soft phonon mode along the $\Gamma$--K path for reduced \textit{tsmear} values, as shown in Fig.~\ref{fig8}(a). The occurrence of this phonon softening is in line with our findings for bulk-layered Nb$_2$CS$_2$ (see Fig.~\ref{fig5}(a)). The phonon softening in bulk-layered Nb$_3$C$_2$S$_2$ disappears for higher \textit{tsmear} values (e.g. 0.0050 Ha), as seen in Fig.~\ref{fig8}(a). The obtained \textit{e-ph} coupling constant, depicted in Fig.~\ref{fig8}(b), is more than double those of the other considered functionalized niobium-carbide MXenes. The strong electron-phonon coupling, boosted by the nearly localized Nb $d$ states around the Fermi level, shown in Fig.~\ref{fig8}(c), gives rise to an elevated $T_c$ of $\sim30$ K. 

The effect of dimensional reduction (exfoliation) on the properties of Nb$_3$C$_2$S$_2$ is similar to the Nb$_2$C\textit{T}$_2$ cases. The calculated phonon dispersion for \textit{tsmear} = 0.0050  Ha, and the corresponding \textit{e-ph} coupling properties remain akin to the bulk-layered case, as shown in the Supplementary Material, resulting in an unchanged $T_c$ of 30 K for monolayer Nb$_3$C$_2$S$_2$. 

\section{Conclusions}
In this work, we set out to theoretically identify the effects of selected functionalizations on superconductivity in the otherwise not superconducting Nb$_2$C MXene. Our first-principles calculations of the superconducting properties of Nb$_2$CCl$_2$ in bulk-layered form revealed a good agreement with the recently experimentally measured $T_c$ of 6 K \cite{Kamysbayev}. In addition, we have shown that superconductivity in Nb$_2$CCl$_2$ persists in monolayer form, even with a slightly increased $T_c$ of $\sim$10 K. Furthermore, the occurrence of Nb $d$ states with a flat dispersion just below the Fermi level enables an enhancement of the $T_c$ through gating, of up to 35 K in bulk-layered form, and even 40 K in the monolayer form. 

Our calculations also revealed a reasonable agreement with the experiment for bulk-layered Nb$_2$CS$_2$. However, the present phonon softening signals possible instability of this crystal in the pure layered structure with $P6_3/mmc$ symmetry. Further analysis including anharmonic phonon effects may elucidate the influence of a possible lattice reconstruction on the superconducting properties of this crystal. On the other hand, our analysis of the monolayer form of Nb$_2$CS$_2$ shows that the phonon softening practically disappears, and that the expected $T_c$ (12 K) is almost twice as high as the measured $T_c$ of its bulk counterpart ($\sim6.5$ K) \cite{Kamysbayev}. 

Our calculations for functionalized Nb$_3$C$_2$ crystals yielded quite surprising results. First of all, we find that Cl functionalization renders this crystal dynamically unstable. Contrarily, both bulk-layered and two-dimensional Nb$_3$C$_2$S$_2$ crystals are stable (albeit showing a similar phonon softening as in bulk Nb$_2$CS$_2$) and superconducting with elevated $T_c$ values of $\sim$30 K. 

Overall, our extensive first-principles exploration clearly demonstrates the potential of surface functionalization to induce superconductivity in MXenes which are not superconducting in pristine form, with critical temperatures that can be strongly enhanced through gating, owing to the presence of quasi-localized electronic states. With further stacking degrees of freedom, as well as possibilities for tailoring the positions of functional atoms and groups inside an extended stack, our findings support the promise of engineered functionalization towards robust superconductivity in MXenes.

\section*{Acknowledgements}
This work is supported by Research Foundation-Flanders (FWO), The Scientific and Technological Research Council of Turkey (TUBITAK) under the contract number COST-118F187, the Air Force Office of Scientific Research under award number FA9550-19-1-7048, and the EU-COST Action CA21144 SUPERQUMAP. Computational resources were provided by the High Performance and Grid Computing Center (TRGrid e-Infrastructure) of TUBITAK ULAKBIM and by the VSC (Flemish Supercomputer Center), funded by the FWO and the Flemish Government -- department EWI. J.B. is a senior postdoctoral fellow of the FWO.

\section*{References}
\bibliographystyle{iopart-num}

\begin{thebibliography}{10}
\expandafter\ifx\csname url\endcsname\relax
  \def\url#1{{\tt #1}}\fi
\expandafter\ifx\csname urlprefix\endcsname\relax\def\urlprefix{URL }\fi
\providecommand{\eprint}[2][]{\url{#2}}

\bibitem{super_cap}
Xia Y, Mathis T~S, Zhao M~Q, Anasori B, Dang A, Zhou Z, Cho H, Gogotsi Y and
  Yang S 2018 {\em Nature\/} {\bf 557} 409--412

\bibitem{battery}
Zhang C~J, Park S~H, Seral‐Ascaso A, Barwich S, McEvoy N, Boland C~S, Coleman
  J~N, Gogotsi Y and Nicolosi V 2019 {\em Nature Communications\/} {\bf 10} 849

\bibitem{battery2}
Siriwardane E~M~D, Demiroglu I, Sevik C, Peeters F~M and Çakır D 2020 {\em
  The Journal of Physical Chemistry C\/} {\bf 124} 21293--21304

\bibitem{Yorulmaz_2020}
Yorulmaz U, Demiro{\u{g}}lu {\.{I}}, {\c{C}}akir D, Gülseren O and Sevik C
  2020 {\em Journal of Physics: Energy\/} {\bf 2} 032006

\bibitem{EMIS}
Shahzad F, Alhabeb M, Hatter C~B, Anasori B, Hong S~M, Koo C~M and Gogotsi Y
  2016 {\em Science\/} {\bf 353} 1137--1140

\bibitem{EMIS2}
Kandemir Z, Torun E, Paleari F, Yelgel C and Sevik C 2022 {\em Phys. Rev.
  Materials\/} {\bf 6}(2) 026001

\bibitem{app1}
Guo J, Legum B, Anasori B, Wang K, Lelyukh P, Gogotsi Y and Randall C~A 2018
  {\em Advanced Materials\/} {\bf 30} 1801846

\bibitem{ReviewMX}
{Dadashi Firouzjaei} M, Karimiziarani M, Moradkhani H, Elliott M and Anasori B
  2022 {\em Materials Today Advances\/} {\bf 13} 100202 ISSN 2590-0498

\bibitem{ReviewMX2}
Gogotsi Y and Anasori B 2019 {\em ACS Nano\/} {\bf 13} 8491--8494

\bibitem{PRM2074002}
Khazaei M, Wang V, Sevik C, Ranjbar A, Arai M and Yunoki S 2018 {\em Phys. Rev.
  Materials\/} {\bf 2}(7) 074002

\bibitem{ReviewMX3}
Naguib M, Barsoum M~W and Gogotsi Y 2021 {\em Advanced Materials\/} {\bf 33}
  2103393

\bibitem{Surface}
Ibragimova R, Erhart P, Rinke P and Komsa H~P 2021 {\em The Journal of Physical
  Chemistry Letters\/} {\bf 12} 2377--2384

\bibitem{Surface2}
Wang C, Chen S and Song L 2020 {\em Advanced Functional Materials\/} {\bf 30}
  2000869

\bibitem{Kamysbayev}
Kamysbayev V, Filatov A~S, Hu H, Rui X, Lagunas F, Wang D, Klie R~F and Talapin
  D~V 2020 {\em Science\/} {\bf 369} 979--983

\bibitem{WANG2022101711}
Wang K, Jin H, Li H, Mao Z, Tang L, Huang D, Liao J~H and Zhang J 2022 {\em
  Surfaces and Interfaces\/} {\bf 29} 101711 ISSN 2468-0230

\bibitem{D0NR03875J}
Bekaert J, Sevik C and Milošević M~V 2020 {\em Nanoscale\/} {\bf 12}(33)
  17354--17361

\bibitem{D2NR01939F}
Bekaert J, Sevik C and Milošević M~V 2022 {\em Nanoscale\/} {\bf 14}(27)
  9918--9924

\bibitem{nbc2dfpt}
Wang S~Y, Pan C, Tang H, Wu H~Y, Shi G~Y, Cao K, Jiang H, Su Y~H, Zhang C, Ho
  K~M and Wang C~Z 2022 {\em The Journal of Physical Chemistry C\/} {\bf 126}
  3727--3735

\bibitem{Gonze2020}
Gonze X, Amadon B, Antonius G, Arnardi F, Baguet L, Beuken J~M, Bieder J,
  Bottin F, Bouchet J, Bousquet E, Brouwer N, Bruneval F, Brunin G, Cavignac T,
  Charraud J~B, Chen W, Côté M, Cottenier S, Denier J, Geneste G, Ghosez P,
  Giantomassi M, Gillet Y, Gingras O, Hamann D~R, Hautier G, He X, Helbig N,
  Holzwarth N, Jia Y, Jollet F, Lafargue-Dit-Hauret W, Lejaeghere K, Marques
  M~A~L, Martin A, Martins C, Miranda H~P~C, Naccarato F, Persson K, Petretto
  G, Planes V, Pouillon Y, Prokhorenko S, Ricci F, Rignanese G~M, Romero A~H,
  Schmitt M~M, Torrent M, van Setten M~J, Troeye B~V, Verstraete M~J, Zérah G
  and Zwanziger J~W 2020 {\em Comput. Phys. Commun.\/} {\bf 248} 107042

\bibitem{Gonze2016}
Gonze X, Jollet F, Abreu~Araujo F, Adams D, Amadon B, Applencourt T, Audouze C,
  Beuken J~M, Bieder J, Bokhanchuk A, Bousquet E, Bruneval F, Caliste D, Côté
  M, Dahm F, Da~Pieve F, Delaveau M, Di~Gennaro M, Dorado B, Espejo C, Geneste
  G, Genovese L, Gerossier A, Giantomassi M, Gillet Y, Hamann D, He L, Jomard
  G, Laflamme~Janssen J, Le~Roux S, Levitt A, Lherbier A, Liu F, Lukačević I,
  Martin A, Martins C, Oliveira M, Poncé S, Pouillon Y, Rangel T, Rignanese
  G~M, Romero A, Rousseau B, Rubel O, Shukri A, Stankovski M, Torrent M,
  Van~Setten M, Van~Troeye B, Verstraete M, Waroquiers D, Wiktor J, Xu B, Zhou
  A and Zwanziger J 2016 {\em Comput. Phys. Commun.\/} {\bf 205} 106--131 ISSN
  0010-4655

\bibitem{PBE}
Perdew J~P, Burke K and Ernzerhof M 1996 {\em Phys. Rev. Lett.\/} {\bf 77}(18)
  3865--3868

\bibitem{HGH}
Hartwigsen C, Goedecker S and Hutter J 1998 {\em Phys. Rev. B\/} {\bf 58}(7)
  3641--3662

\bibitem{DFPT}
Savrasov S~Y and Savrasov D~Y 1996 {\em Phys. Rev. B\/} {\bf 54}(23)
  16487--16501

\bibitem{Eliashberg1}
Eliashberg G~M 1960 {\em J. Exp. Theor.\/} {\bf 11}(3) 696

\bibitem{Eliashberg2}
Eliashberg G~M 1961 {\em J. Exp. Theor.\/} {\bf 12}(5) 1000

\bibitem{Eliashberg3}
Giustino F 2017 {\em Rev. Mod. Phys.\/} {\bf 89}(1) 015003

\bibitem{PhysRev.167.331}
McMillan W~L 1968 {\em Phys. Rev.\/} {\bf 167}(2) 331--344

\bibitem{PhysRevB.12.905}
Allen P~B and Dynes R~C 1975 {\em Phys. Rev. B\/} {\bf 12}(3) 905--922

\bibitem{ALLEN19831}
Allen P~B and Mitrović B 1983 {\em Solid State Phys.\/} {\bf 37} 1--92 ISSN
  0081-1947

\bibitem{Grimvall}
Grimvall G 1981 {\em The electron-phonon interaction\/} (North Holland
  Publishing Co. (Amsterdam))

\bibitem{Grimme2006}
Grimme S 2006 {\em J. Comput. Chem.\/} {\bf 27} 1787--1799 ISSN 0192-8651,
  1096-987X

\bibitem{Grimme2010}
Grimme S, Antony J, Ehrlich S and Krieg H 2010 {\em The Journal of Chemical
  Physics\/} {\bf 132} 154104 ISSN 0021-9606, 1089-7690

\bibitem{Becke2006}
Becke A~D and Johnson E~R 2006 {\em The Journal of Chemical Physics\/} {\bf
  124} 221101

\bibitem{PhysRevLett.125.106101}
Bianco R, Monacelli L, Calandra M, Mauri F and Errea I 2020 {\em Phys. Rev.
  Lett.\/} {\bf 125}(10) 106101
\end{thebibliography}
\providecommand{\noopsort}[1]{}\providecommand{\singleletter}[1]{#1}%
\providecommand{\newblock}{}

\end{document}